\documentstyle[12pt]{ioplppt}
\eqnobysec

\begin{document}
\jl{1}

\title{Surface critical exponents for a three-dimensional
modified spherical model}[Surface critical exponents]
\author{D M Danchev, J G Brankov and M E Amin}
\address{Institute of Mechanics, Bulgarian Academy of Sciences,\\
Acad. G. Bonchev Str., block 4, 1113 Sofia, Bulgaria}

\begin{abstract}
A modified three-dimensional mean spherical model with a $L$-layer film
geometry under Neumann-Neumann boundary conditions is considered. Two
spherical fields are present in the model: a surface one fixes the mean
square value of the spins at the boundaries at some $\rho > 0$, and a bulk
one imposes the standard spherical constraint (the mean square value of the
spins in the bulk  equals one). The surface susceptibility $\chi_{1,1}$ has
been evaluated exactly. For $\rho =1$ we find that $\chi_{1,1}$ is finite
at
the bulk critical temperature $T_c$, in contrast with the recently derived
value $\gamma_{1,1}=1$ in the case of just one global spherical constraint.
The result $\gamma_{1,1}=1$ is recovered only if $\rho =\rho_c= 2-(12 K_c
)^{-1}$, where $K_c$ is the dimensionless critical coupling. When $\rho >
\rho_c$, $\chi_{1,1}$ diverges exponentially as $T\rightarrow T_c^{+}$. An
effective hamiltonian which leads to an exactly solvable model with
$\gamma_{1,1}=2$, the value for the $n\rightarrow \infty $ limit of the
corresponding $O(n)$ model, is proposed too.
\end{abstract}
\pacs{05.20.-y, 05.50.+q, 75.10.Hk}
\maketitle

\section{Introduction}

In a recent article \cite{DBM97} (hereafter referred to as I) the
finite-size scaling behaviour of a three-dimensional system with a film
geometry $L\times \infty ^2$ has been investigated within the mean
spherical
model with Neumann-Neumann and Neumann-Dirichlet boundary conditions and
surface fields $h_1$ and $h_{_L}$ acting at the boundaries. The obtained
results imply the well known exponent $\Delta _1^{{\rm o}}=1/2$ for the
ordinary surface phase transition at a Dirichlet boundary, and the
emergence
of a new critical exponent $\Delta _1^{{\rm sb}}=3/2$, characterizing the
Neumann boundary (for a general review on surface critical phenomena see,
e.g. \cite{B83} - \cite{Di90}, and on finite-size scaling \cite{B83},
\cite{Ba83} - \cite{krech}). The conjecture has been made that the latter
critical exponent corresponds to the special (surface-bulk) phase
transition
within the model. The last is in consistence with the general expectation
for the finite-size scaling form of the free energy for this type of phase
transitions if one assumes that the crossover exponent $\Phi =0$, as it is
for three-dimensional $O(n)$ models \cite{B83}. It has also been derived
that the critical exponent of the local surface susceptibility $\chi
_{1,1}$
is $\gamma _{1,1}^{{\rm sb}}=1$. The same result is known to hold for the
spherical model with enhanced surface couplings under Dirichlet-Dirichlet
boundary conditions \cite{BJSW}. Unfortunately, in that case the model
quite
unphysically predicts that the surface orders for a sufficiently large
enhancement at some temperature above the bulk critical one even for $d=3$.
This is no more the case when one improves the model by introducing a
second
spherical constraint on the spins at the boundaries \cite{SJB}, since the
only critical point that remains for $d\leq 3$ is the bulk one. Then for
$d=3
$ the exponent $\gamma _{1,1}^{{\rm o}}=-1$ corresponds to an ordinary
phase
transition \cite{B83,Ba74}. In I the case of equal bulk and surface
couplings was considered and the question if and how the surface behaviour
of the system with Neumann-Neumann boundary conditions will change under
additional spherical constraints on the spins at and near the surfaces was
left open. One of the aims of the present article is to contribute in
clarifying that point. To this end we consider the critical behaviour of
the
local surface susceptibility
\begin{equation}
\chi _{1,1}(T;\rho )=\lim_{L\rightarrow \infty }\left[ -L\partial
^2f_L ( T,h_1,h_L;\rho ) /\partial h_1^2\right]_{\mid h_1=h_L=0}
\label{lss}
\end{equation}
in the case when the mean square value of the spins at the boundaries is
fixed at some positive number $\rho $ by an additional spherical
constraint.
The model defined in this way will be termed modified spherical model. In
equation (\ref{lss}) and in the remainder $f_L ( T,h_1,h_L;\rho )$
denotes the free energy density (per $k_B T$ and per spin) of a
three-dimensional hypercubic lattice system with a film geometry $L\times
\infty ^2$ at temperature $T$. Neumann-Neumann boundary conditions are
imposed across the finite dimension of extent $L$. Surface fields $h_1$ and
$h_{_L}$ are supposed to act at the surfaces bounding the system. Since we
are interested in the case $h_1=h_L$, only one additional constraint on the
boundary spins is imposed. It turns out that the behaviour of $\chi _{1,1}$
in the vicinity of the bulk critical point $T_c$ depends crucially on $\rho
$%
. It will be shown that only if $\rho =\rho _c:=2-(12K_c)^{-1}=1.34053\dots
$
one obtains the previously found value $\gamma _{1,1}=\gamma _{1,1}^{{\rm
sb}%
}=1.$ If $\rho <\rho _c$ then $\chi _{1,1}$ has the singularity
characteristic of the spherical model with Dirichlet boundary conditions,
i.e. $\gamma _{1,1}=\gamma _{1,1}^{{\rm o}}=-1$. When $\rho >\rho _c$ then
$%
\chi _{1,1}$ diverges exponentially as $T\rightarrow T_c^{+}$, which
reminds
the behaviour of a two-dimensional $O(n)$, $n>2$, model close to $T=0$. The
calculation of the mean square value $\rho _s$ of the spins at a Neumann
boundary in the standard spherical model elucidates the appearance of the
critical value of $\rho =\rho _c$: it turns out that $\rho _c=\rho _s$.
Moreover, we note that the second spherical field, to be denoted by $v$
(see
equation (\ref{H}) below), can be considered as a free parameter. It will
be
shown that changing it one interpolates continuously from Neumann, via
mixed,
to Dirichlet boundary conditions. Following I, under Neumann boundary
conditions we mean here the case when the interaction of the finite system
with the ''environment'' is modelled by letting the spins surrounding the
system take the same values as their nearest neighbour inside the system.
Under Dirichlet boundary conditions this interaction is modelled by fixing
to zero value the spin configuration outside the system. (For a precise
mathematical definition of the boundary conditions see I.) The mixed
boundary conditions correspond then to the situation when the spins
surrounding the system are set to take values proportional (but not equal)
to those of their nearest neighbour inside the system. Obviously, the above
terminology is justified by analogy with the continuum limit. Note that for
any $v$, just due to the symmetry which arises from the identical boundary
conditions and fields ($h_1=h_L$) at the opposite surfaces, the system
models by itself an analogue of a Neumann boundary at the middle layers.
Therefore, if one considers the local surface succeptibility $\chi _{l,l}$
for the $l$-th layer, one would expect to obtain the critical exponent for
the Neumann boundary, $\gamma _{l,l}=1$, for $l$ around the middle of the
system. The last is obviously true even if the system is with otherwise
Dirichlet boundary conditions. Finally, it will be shown that if $v$ is a
given function of the temperature, one obtains $\gamma _{1,1}^{{\rm
sb}}=2$,
which is the corresponding value for the $O(n)$ model in the limit $%
n\rightarrow \infty $.

As it is well known, the infinite translational invariant spherical model
is
equivalent to the $n\rightarrow \infty $ limit of a similar system of $n$%
-component vectors \cite{S68},\cite{KT71}. However, the spherical model
with
surfaces (or, more generally, without translation-invariant symmetry) is in
fact {\em not} such a limit \cite{K73} (for the results available for the
spherical model see, e.g.,\cite{B83}, \cite{Ba83}, \cite{P90}, \cite{krech}
and references therein). In other words, the spherical model under
nonperiodic boundary conditions is not in the same surface universality
class as the corresponding $O(n)$ model in the limit $n\rightarrow \infty
$,
in contrast with the bulk universality classes. The last becomes apparent
when one investigates surface phase transitions for an $O(n)$ model in the
limit $n\rightarrow \infty $. In that case one obtains \cite{B83} $\Delta
_1=1/(d-2)$ (i.e. $\Delta _1=1$ for $d=3$) for ordinary and $\Delta
_1=2/(d-2)$ (i.e. $\Delta _1=2$ for $d=3$) for special phase transitions.
It
is believed that the corresponding equivalence will be recovered if one
imposes spherical constraints in a way which ensures that the mean square
value of each spin of the system is the same \cite{K73} (unfortunately
such a model is rather untractable). One of the goals of the present work
is
to see if, and up to what extent, the behaviour of the system with two
spherical fields will be closer to the $O(n)$ model in the limit $%
n\rightarrow \infty $, in comparison with the standard mean spherical model
(with only one spherical field).

The article is organized as follows. In Section 2 we describe the model and
present convenient starting expressions for the mean spherical constraints,
the free energy density and the local surface susceptibility. Our main
results on the behaviour of $\chi _{1,1}$ and $\chi _{l,l}$ as a function
on
$T$ and $\rho $ are given in Section 3. The paper closes with a short
discussion given in Section 4.

\section{The model}

We consider explicitly the three dimensional mean spherical model with
nearest-neighbour ferromagnetic interactions on a simple cubic lattice. At
each lattice site $\vec r=(r_1,r_2,r_3)\in {\bf Z}^3$ there is a random
(spin) variable $\sigma (\vec r)\in {\bf R}$ and the energy of a
configuration $\sigma _{_\Lambda }=\{\sigma (\vec r),\vec r\in \Lambda \}$
in a finite domain $\Lambda \subset {\bf Z}^3,\Lambda =L_1\times L_2\times
L_3$, containing $|\Lambda |$ sites, is given by
\begin{equation}
\fl \beta {\cal H}_{_\Lambda }^{(\tau )}(\sigma _{_\Lambda }|K,h_{_\Lambda
};s)=-K\sigma _{_\Lambda }^{\dag }\cdot Q_{_\Lambda }^{(\tau )}\cdot \sigma
_{_\Lambda }+s\sigma _{_\Lambda }^{\dag }\cdot \sigma _{_\Lambda }+v\sigma
_s^{\dag }\cdot \sigma _s-h_{_\Lambda }^{\dag }\cdot \sigma _{_\Lambda }.
\label{H}
\end{equation}
Here the $|\Lambda |\times |\Lambda |$ interaction matrix $Q_{_\Lambda
}^{(\tau )}$ can be written as
\begin{equation}
Q_{_\Lambda }^{(\tau )}=(\Delta _1^{(\tau _1)}+2E_1)\times (\Delta
_2^{(\tau
_2)}+2E_2)\times (\Delta _3^{(\tau _3)}+2E_3),  \label{Q}
\end{equation}
where $\times $ denotes the outer product of the corresponding matrices, $%
\Delta _i^{(\tau _i)}$ is the $L_i\times L_i$ discrete Laplacian under
boundary conditions $\tau _i$, and $E_i$ is the $L_i\times L_i$ unit
matrix.
In equation (\ref{H}) $\beta =1/k_BT$ is the inverse temperature; $K=\beta
J$
is the dimensionless coupling constant; $h_{_\Lambda }=\{h(\vec r),\vec
r\in
\Lambda \}$, with $h(\vec r)\in {\bf R}$, is an external magnetic field;
$s$
and $v$ are the spherical fields which are to be determined from the mean
spherical constraints (see below); $\sigma_s=\{\sigma(\vec r),\vec r\in
S\}$%
, $S=\{ (r_1,r_2,1)\cup (r_1,r_2,L_3) \}$, $r_1=1,\ldots,L_1$, $%
r_2=1,\ldots,L_2$.

The free-energy density of the modified mean shperical model in a finite
region $\Lambda $ is given by the Legendre transformation
\begin{equation}
\fl \beta f_{_\Lambda }^{(\tau )}(K,h_{_\Lambda };\rho ):=\sup_{s,v}\left\{
-|\Lambda |^{-1}\ln Z_{_\Lambda }^{(\tau )}(K,h_{_\Lambda };s,v)-s-\rho
v|S|/|\Lambda |\right\} ,  \label{F}
\end{equation}
where $|S|$ is the total number of spins at the boundaries $S$ and
\begin{equation}
Z_{_\Lambda }^{(\tau )}(K,h_{_\Lambda };s,v)=\int_{{\bf R}^{|\Lambda
|}}\exp
\left[ -\beta {\cal H}_{_\Lambda }^{(\tau )}(\sigma_{_\Lambda
}|K,h_{_\Lambda };s,v)\right] \prod_{\vec r\in \Lambda } \d \sigma (\vec r)
\label{Z}
\end{equation}
is the partition function. The supremum is attained at the solutions of the
mean spherical constraints
\begin{equation}
\langle \sigma_{_\Lambda }^{\dag }\cdot \sigma_{_\Lambda }\rangle =|\Lambda
|
\label{s}
\end{equation}
and
\begin{equation}
\langle \sigma_s^{\dag }\cdot \sigma_s\rangle =\rho |S|,  \label{v}
\end{equation}
where $\langle \cdots \rangle $ denotes expectation value with respect to
the hamiltonian $\beta {\cal H}_{_\Lambda }^{(\tau )}(\sigma _{_\Lambda
}|K,h_{_\Lambda };s)$. Let us denote by $-2+2\cos \varphi _{L_i}^{\tau
_i}(k_i)$, $k_i=1,\dots ,L_i,$ $i=1,2,3$, the eigenvalues of the matrix $%
\Delta_i^{(\tau _i)}$ . Let us further suppose, say, periodic boundary
conditions across $L_1$ and $L_2$ and Neumann-Neumann boundary conditions
across $L_3$. Then, by direct evaluation of the integrals in the partition
function (\ref{Z}), after taking the limit $L_1,L_2\rightarrow \infty $ at
a
fixed $L_3=L$, one obtains for the free energy
\begin{eqnarray}
\fl \beta f_L^{(n)}(K,h_{_1},h_L;\rho ) &=&\frac{1}{2}\log \frac{K}{\pi}
-6K+
\nonumber \\
&&\ \sup_{\phi ,\omega }\left\{ \frac{1}{L}\frac{1}{8\pi ^2}\int_0^{2\pi}
\d \theta_1\int_0^{2\pi }\d \theta_2\sum_{k=1}^{L}\log [ \phi
+2\sum_{i=1}^2(1-\cos \theta_i)]+ \right.  \nonumber \\
&&\ 2[1-\cos \varphi_L^n(k;\omega )] -\frac{1}{4KL}\sum_{k=1}^{L}\frac{%
\left| h_L^{(n)}(k;\omega )\right|^2} {\phi +2[1-\cos \varphi_L^n(k;\omega
)]} - \nonumber \\
&&\ \left. K\left( \phi +\frac{2}{L}\rho \omega \right) \right\} .
\label{FF}
\end{eqnarray}
Here $-2+2\cos \varphi_L^n(k;\omega )$, $k=1,\dots ,L$ are the eigenvalues
of the matrix
\begin{equation}
\Delta _L^{(n)}(\omega )=\Delta _L^{(n)}-\omega ( \delta _{1,1}+\delta
_{L,L}) ,  \label{D}
\end{equation}
and
\begin{equation}
h_L^{(n)}(k;\omega )=h_{_1}u_{_L}^n(1,k;\omega )+h_Lu_{_L}^n(L,k;\omega )
\label{hm}
\end{equation}
where $\{u_{_L}^n(r,k;\omega ),r=1,\ldots ,L\},$ $k=1,\ldots ,L$ are its
eigenvectors; the superscript $n$ stays there for Neumann-Neumann boundary
conditions. In equations (\ref{FF}) - (\ref{hm}) use has been made of the
following definitions
\begin{equation}
\phi =s/K-6,  \label{fi}
\end{equation}
and
\begin{equation}
\omega =v/K.  \label{omega}
\end{equation}
From the requirement for existence of the partition function one has the
constraint
\begin{equation}
\phi +2\min_{k=1,\dots, L}[1-\cos \varphi_L^n(k;\omega )] >0.  \label{C}
\end{equation}
The eigenvalues and the eigenvectors of the matrix $\Delta _L^{(n)}(\omega
)$
can be obtained in a way simmilar to the one used in \cite{SJB} and \cite
{Z92}. The results are:

i) For given $L$ and $\omega $, when $\left| 1-\omega \right| \neq 1$, the
numbers $\varphi _L^n(k;\omega )$, $k=1,\dots ,L$ are the $L$ roots of the
equations
\begin{equation}
1-\omega =\frac{\sin \left[ \frac{1}{2}(L+1) \varphi \right]}{\sin
\left[\frac{1}{2}(L-1) \varphi \right] }  \label{sin}
\end{equation}
and
\begin{equation}
1-\omega =\frac{\cos \left[\frac{1}{2}(L+1) \varphi \right]}{\cos
\left[\frac{1}{2}(L-1) \varphi \right]}  \label{cos}
\end{equation}
with $0<$Re$(\varphi )<\pi $ and Im$(\varphi )>0$. For concreteness and
simplification of the notations below, without loss of generality in the
final results, we assume $L$ to be an odd integer. Then, it is easy to see
that equation (\ref{sin}) possesses $(L-1)/2$ solutions of the specified
type, whereas equation (\ref{cos}) gives the remaining $(L+1)/2$ solutions.
From (\ref{sin}) and (\ref{cos}) one obtains that (for fixed $L$ and $k$ )
\begin{equation}
\frac{d\varphi }{d\omega }=\frac 1L\frac{2\sin \varphi }{(1-\omega
)^2-2(1-\omega )\cos \varphi +1+L^{-1}[1-(1-\omega )^2]}.  \label{Der}
\end{equation}
Further, if $\left| 1-\omega \right| <1$ all the $L$ roots are real. In
that
case there is only one root of equation (\ref{sin}) per interval $(2\pi
k/(L-1),2\pi (k+1)/(L-1))$, $k=0,\dots ,(L-3)/2$. Similarly, equation (\ref
{cos}) has only one root per interval $(\pi (2k-1)/(L-1),\pi
(2k+1)/(L-1))$,
$k=1,\dots ,(L-3)/2$, and one root in each of the intervals $(0,\pi
/(L-1))$
and $(\pi -\pi /(L-1),\pi )$. Let us consider now the case $|1-\omega |
>1$.
Then, if $\omega <0$, one again has only one root of equation
(\ref{sin}) per interval $(2\pi k/(L-1),2\pi (k+1)/(L-1))$, $k=1,\dots
,(L-3)/2$, and, similarly, one root of equation (\ref{cos}) per interval
$(\pi (2k-1)/(L-1),\pi (2k+1)/(L-1))$, $k=1,\dots ,(L-3)/2$, and one root
in
$(\pi -\pi /(L-1),\pi )$, i.e. altogether $L-2$ real roots in the interval
$(0,\pi )$. The remaining two roots are given by
\begin{equation}
\varphi _0=i\log (1-\omega )\pm O\left( (1-\omega )^{-(L-1)}\right) .
\label{cr}
\end{equation}
(Strictly speaking the roots are degenerate only up to exponentially small
corrections.) In the case $\omega >2$ one again has $L-2$ real roots in the
interval $\left( 0,\pi \right) $ and the remaining two roots are then given
by $\varphi _0=\pi +i\log (\omega -1)\pm O\left( (\omega
-1)^{-(L-1)}\right)
$.

ii) The components of the eigenvectors $\{u_{_L}^n(r,k;\omega ),r=1,\ldots
,L\},$ $k=1,\ldots ,L$ of the matrix $\Delta _L^{(n)}(\omega )$ are given
by
the expression ($|1-\omega |\neq 1$)
\begin{equation}
\fl u_{_L}^n(r,k;\omega )=\sqrt{\frac 2L}\frac{\sin [r\varphi _L^n(k;\omega
)]-(1-\omega )\sin [(r-1)\varphi _L^n(k;\omega )]}{\{(1-\omega
)^2-2(1-\omega )\cos \varphi _L^n(k;\omega )+1+L^{-1}[1-(1-\omega
)^2]\}^{1/2}}.  \label{ukw}
\end{equation}

iii) For completenes we give also the results for the well known case $%
\left| 1-\omega \right| =1$ (see, e.g, I, \cite{Z92}). Then $\varphi
_L^n(k;0)=\pi (k-1)/L$, $\varphi _L^n(k;2)=\pi k/L$, $k=1,\dots ,L$, and
the
components of the eigenvectors are $u_{_L}^n(r,k;0)=$ $\sqrt{(2-\delta
_{k,1})/L}\cos[(r-1/2)\varphi _L^n(k;0)]$ and $u_{_L}^n(r,k;2)=$
$\sqrt{2/L}%
\sin[(r-1/2)\varphi _L^n(k;2)]$, respectively.

Finally, we remind that we are interested mainly in the behaviour of the
local surface susceptibility for which we obtain from equations (\ref{lss})
and (\ref{FF})
\begin{equation}
\chi _{1,1}(T;\rho )=\frac 1{2K}\lim_{L\rightarrow \infty
}\sum_{k=1}^L\frac{%
| u_L^n(1,k;\omega )|^2}{\phi +2[1-\cos \varphi _L^n(k;\omega )]}.
\label{hid}
\end{equation}
If, instead of the local susceptibility at the surface of the system, one
is
interested in the local susceptibility of the $l$-th layer, $\chi
_{l,l}(T;\rho )$, the corresponding result reads
\begin{equation}
\chi _{l,l}(T;\rho )=\frac 1{2K}\lim_{L\rightarrow \infty
}\sum_{k=1}^L\frac{%
| u_L^n(l,k;\omega )| ^2}{\phi +2[1-\cos \varphi _L^n(k;\omega )]}
\label{hild}
\end{equation}
The above expression can be obtained in a way analogous to the derivation
of
$\chi _{1,1}$ by imposing a local magnetic field $h_l$ on the spins in the
$l$-th layer.

To determine the behaviour of the spherical fields $\phi $ and $\omega $
one
has to analyze equations (\ref{s}) and (\ref{v}). From (\ref{F}),
(\ref{FF}%
), (\ref{fi}), (\ref{omega}) and (\ref{Der}) one obtains explicitly the set
of equations
\begin{equation}
\fl 2K=\frac{1}{4\pi^2}\int_0^{2\pi }\d \theta_1\int_0^{2\pi }\d \theta_2
\frac{1}{L}\sum_{k=1}^L\left\{\phi +2\sum_{i=1}^2(1-\cos \theta_i)+2[1-\cos
\varphi_L^n(k;\omega )]\right\}^{-1}  \label{efi}
\end{equation}
and
\begin{eqnarray}
\fl
2K\rho =\frac{1}{4\pi^2}\int_0^{2\pi }\d \theta _1\int_0^{2\pi }\d \theta_2
\frac{1}{L}\sum_{k=1}^L2\sin^2\varphi_L^n(k;\omega )\times   \nonumber \\
\left\{ \phi +2\sum_{i=1}^2(1-\cos \theta _i)+2[1-\cos \varphi
_L^n(k;\omega )]\right\}^{-1}\times   \nonumber \\
\left\{ (1-\omega )^2-2(1-\omega )\cos \varphi _L^n(k;\omega )+1+
L^{-1}[1-(1-\omega )^2]\right\}^{-1}.  \label{eom}
\end{eqnarray}
These equations determine the point at which the finite-size free energy
density (\ref{FF}), which is analytical and strictly concave function of
$\phi $ and $\omega $ in the domain given by inequality (\ref{C}), reaches
its global maximum. Clearly, in the thermodynamic limit the free energy
density
is independant of the surface spherical field $\omega $. As it is well
known,
for all $K\geq K_c$ its supremum sticks to the endpoint $\phi _0=0$ of the
allowed interval $\phi_0>0$, where the bulk free energy density is finite.
When $K<K_c$, the supremum is attained at a point $\phi _0=\phi _0(K)>0$,
which satisfies the limit form of equation (\ref{efi}) \cite{BF},
\begin{equation}
2K=W_3(\phi _0),  \label{beq}
\end{equation}
where
\begin{equation}
W_d(\phi )=\frac 1{\pi^d}\int_0^\pi \d\theta_1\cdots \int_0^\pi \d \theta_d
\left[ \phi +2\sum_{i=1}^d(1-\cos \theta_i)\right]   \label{W}
\end{equation}
is the $d$-dimensional Watson integral, $W_d(0)=2K_c$. In general, the
solutions $\phi $ and $\omega $ of equations (\ref{efi}) and (\ref{eom})
can
be written in the form $\phi =\phi _0+\Delta \phi ,\omega =\omega _0+\Delta
\omega ,$ where $\Delta \phi $ and $\Delta \omega $ tend to zero when
$L\rightarrow \infty ,\ $and $\phi _0$ and $\omega _0$ are solutions of the
corresponding equations where the limit $L\rightarrow \infty $ is taken.

Equations (\ref{cr}) - (\ref{W}) provide the basis for our further
analysis. Before passing to it we note that, instead of considering $\omega
$
as a variable that has to be determined from equations (\ref{efi}) and
(\ref
{eom}), one can consider it as an additional free parameter. Then, from
equation (\ref{D}) it is clear that $\omega =0$ yields the standard
spherical model with Neumann-Neumann boundary conditions, whereas $\omega
=1$
yields the same model under Dirichlet-Dirichlet boundary conditions. When
$0<\omega <1$ we have mixed (or ''intermediate'' \cite{GF90}) boundary
conditions which interpolate between the above two extreme cases.
Therefore,
in this way one should reproduce the previously known results for the
properties of the local susceptibilities. In addition, as we shall see
later,
by choosing $\omega $ to be a given function of the temperature, one can
define an effective spherical model with $\gamma _{1,1}=2,$ which
corresponds to the critical exponent for the surface-bulk phase transition
within the $O(n)$ model in the limit $n\rightarrow \infty $.

\section{Critical behaviour of the local susceptibilities}

We study here the critical behaviour of the local susceptibilities
$\chi_{1,1}$ and $\chi_{l,l}$ for $l$ close to the middle of the system.

From equations (\ref{hid}) and (\ref{ukw}) we obtain for the surface
susceptibility
\begin{eqnarray}
\fl \chi _{1,1}(T;\rho ) =\frac 1K\lim_{L\rightarrow \infty }\frac
1L\sum_{k=1}^L\frac{\sin ^2\varphi _L^n(k;\omega )}{\phi +2[1-\cos \varphi
_L^n(k;\omega )]}\times   \nonumber \\
\left[ 1-2(1-\omega )\cos \varphi _L^n(k;\omega )+(1-\omega
)^2+[1-(1-\omega )^2]/L\right] ^{-1},  \label{hi1}
\end{eqnarray}
where the limit $L\rightarrow \infty $ in (\ref{hi1}) is to be taken over
the finite-size
solutions $\omega $ and $\phi $ of equations (\ref{efi}) and (\ref{eom}).
If $|1-\omega | <1$, from the properties of $\varphi _L^n(k;\omega )$,
$k=1,\dots ,L$, described in Section 2, it follows that as $L\rightarrow
\infty $ the sum in equation (\ref{hi1}) tends to the corresponding well
defined integral
\begin{equation}
\fl \chi _{1,1}(T;\rho )=\frac 1K\frac 1\pi \int_0^\pi \frac{\sin ^2\varphi
}{\left[ \phi _0+2(1-\cos \varphi )\right] \left[ 1-2(1-\omega _0)\cos
\varphi +(1-\omega _0)^2\right] }d\varphi .  \label{hiI}
\end{equation}
The integral can be taken exactly \cite{GR73} with the result
\begin{equation}
\chi _{1,1}(T;\rho )=\frac 1K\frac 1{\sqrt{\phi _0\left( 4+\phi _0\right)
}%
+\phi _0+2\omega _0}.  \label{hif}
\end{equation}
When $|1-\omega | >1$ one has to take into account the contribution of the
two
complex roots which turns out to be of the same order as the contribution
of
all other roots. The contribution of the latter $L-2$ roots is again given
by
the integral in the r.h.s of equation (\ref{hiI}). Performing the
calculations
one ends up with the same analytical expression for $\chi_{1,1}(T;\rho )$
as
the one given by equation (\ref{hif}).

The surface spherical field $\omega _0$ satisfies the corresponding limit
form of the spherical constraint (\ref{eom}) at fixed $\phi_0=0$ for $K\geq
K_c$, and $\phi _0=\phi _0(K)$ for $K<K_c$. The right-hand side of this
equation can be treated in a way similar to that for (\ref{hi1}). When
$|1-\omega |<1$, due to the properties of the roots $\varphi_L^n(k;\omega
)$,
$k=1,\dots L$, the sum in (\ref{eom}) converges as $L\rightarrow \infty $
to the corresponding well defined integral, which can be taken exactly.
Performing this procedure, one obtains finally
\begin{equation}
2K\rho =G_3(\phi _0,\omega _0),  \label{seq}
\end{equation}
where
\begin{eqnarray}
\fl G_d(\phi ,\omega ) =\frac 2{\pi ^d}\int_0^\pi \d \theta_1\cdots
\int_0^\pi \d \theta_{d-1}\left\{ \phi +2\omega +2\sum_{i=1}^{d-1}(1-\cos
\theta _i) + \right.   \nonumber \\
\left. \left[ \phi +2\sum_{i=1}^{d-1}(1-\cos \theta _i)\right]
^{1/2}\left[ \phi +4+2\sum_{i=1}^{d-1}(1-\cos \theta _i)\right]
^{1/2}\right\} ^{-1}.  \label{G3}
\end{eqnarray}
When $|1-\omega |>1$, one has to treat separately the contribution from the
two complex roots. The contribution from the $L-2$ real roots leads again
to
a well defined integral that can be taken exactly. As for $\chi_{1,1}(T;
\rho )$, the final result is given by the same analytical
expression as in the case $|1-\omega |<1$, i.e. equation (\ref{seq}) is
actually valid for all $\omega _0$ (the restrictions on $\omega _0$ and $%
\phi _0$ stemming from the constraint (\ref{C}) are stated below).

Let us denote by $G_3^{+} ( \phi ,\omega )$ the branch of the function $G_3
(\phi ,\omega)$ defined for $\omega \geq 0$ and by $G_3^{-}(\phi ,\omega)$
the one for $\omega < 0$. Then, by means of identical transformations it is
easy to show that

\begin{equation}
\fl G_3^{-}(\phi ,\omega )=(1-\omega )^{-2}G_3^{+}\left( \phi
,\frac{|\omega
|}{1-\omega }\right) -\frac{\omega (2-\omega )}{(1-\omega )^2}W_2\left(
\phi
-\frac{\omega ^2}{1-\omega }\right) ,  \label{gpm}
\end{equation}
and
\begin{eqnarray}
\fl G_3^{+}(\phi ,\omega ) =(1-\omega )^{-1}\left[ 2W_3(\phi )-\frac
16+\frac 16\phi W_3(\phi )\right] -  \label{gpe} \\
\frac \omega {(1-\omega )(2-\omega )}\frac 2{\pi ^2}\int_0^\pi \d \theta_1
\int_0^\pi \d \theta_2\left\{ \phi +2\sum_{i=1}^2(1-\cos \theta_i)+\right.
\nonumber \\
\left. \frac \omega {2-\omega }\left[ \phi +2\sum_{i=1}^2(1-\cos
\theta _i)\right] ^{1/2}\left[ \phi +4+2\sum_{i=1}^2(1-\cos \theta
_i)\right] ^{1/2}\right\} ^{-1}.  \nonumber
\end{eqnarray}
Finally, in the limit $L\rightarrow \infty $ the constraint (\ref{C}) for
the existence of the partition function yields the allowed domain of values
of the spherical fields,
\begin{equation}
\phi _0\geq \left\{
\begin{array}{ccc}
0, & \text{if} & \omega _0\geq 0 \\
\omega _0^2/\ (1-\omega _0), & \text{if} & \omega _0\leq 0.
\end{array}
\right.   \label{CF}
\end{equation}
From equation (\ref{hif}) it follows that the above inequalities imply
$\chi
_{1,1}(T;\rho )\geq 0$, as it should be expected on general physical
grounds.

As it is evident from equation (\ref{G3}), $G_3(\phi ,\omega )$ is a
monotonically decreasing function of $\omega $ which tends to zero from
above as $\omega \rightarrow +\infty $. Due to inequalities (\ref{CF}), at
$%
\phi =0$ we have to consider it on the half-line $\omega \geq 0$, where it
is bounded from above by its value at $\omega =0$, see equation
(\ref{gpe}),

\begin{equation}
G_3(0,0)=2K_c-1/6:=2K_c\rho _c.  \label{roc}
\end{equation}
On the other hand, if $\phi >0$, the definition domain of $G_3(\phi ,\omega
)
$ is restricted by (\ref{CF}) to the half-line $\omega \geq \omega_1 (\phi
)$,
where
\begin{equation}
\omega_{1}(\phi )= -(\phi +\phi ^2/4)^{1/2}-\phi /2 .  \label{om1}
\end{equation}
From the representation (\ref{gpm}) and the known expansion of $W_2(x)$ as
$%
x\downarrow 0$,
\begin{equation}
W_2(x)=(4\pi )^{-1}\ln x^{-1}+O(1),  \label{w2}
\end{equation}
it follows that $G_3(\phi ,\omega )$ diverges logarithmically to $+\infty $
as $\omega \downarrow \omega _1(\phi )$.

Before passing to the analysis of the above equations, in order to
determine
the behaviour of $\chi _{1,1}(T;\rho )$, let us first consider the simpler
case
of $\omega $ as a free parameter. Then, for Neumann-Neumann boundary
conditions one has (see equations (\ref{Q}) and (\ref{D})) $\omega =0$,
whereas
$\omega =1$ for Dirichlet-Dirichlet boundary conditions. Thus, from
(\ref{hif}) and the well known behavior of $\phi _0$ in the vicinity of the
bulk critical temperature $\phi _0\simeq [ 8\pi (K_c-K)]^2$ \cite{BF}, one
immediately obtains all previously known results for the
critical behaviour of the local surface susceptibility \cite{DBM97},
\cite{Ba74}:

a) Neumann-Neumann boundary conditions ($\omega =0$; the result given below
follows directly from equation (3.5) in \cite{DBM97} for $h_1=h_L$)

\begin{equation}
\chi _{1,1}(T)=\left( 2K\right) ^{-1}\left\{ \phi _0/2+\left[ \phi _0\left(
1+\phi _0/4\right) \right] ^{1/2}\right\} ^{-1},  \label{hiN}
\end{equation}
i.e. $\gamma _{1,1}=1,$ and

b) Dirichlet-Dirichlet boundary conditions ($\omega =1$; see equation
(61) in \cite{Ba74})

\begin{equation}
\chi_{1,1}(T)=\left( 2K\right) ^{-1}\left\{ 1+\phi _0/2+\left[ \phi
_0\left( 1+\phi _0/4\right) \right] ^{1/2}\right\} ^{-1},  \label{hiD}
\end{equation}
i.e. $\gamma_{1,1}=-1$.

For $\omega \neq 0,1$ one has the case of the so-called intermediate \cite
{GF90} boundary conditions. As it is clear from (\ref{hif}), $\chi _{1,1}$
diverges in the vicinity of $T=T_c$ if and only if $\omega =0$, i.e. under
Neumann-Neumann boundary conditions.

Let us now comment on the critical value $\rho_c$ of the parameter $\rho $,
defined in equation (\ref{roc}). By using translation invariance argument,
for the mean square length of the spins at the Newmann boundary of the
standard spherical model with one global spherical field $\phi $ one
obtains
in zero magnetic field
\begin{eqnarray}
\fl
\langle \sigma^2 (r_1 , r_2 , 1 )\rangle =
\frac{1}{2K L_1 L_2} \times \nonumber \\
\sum_{k_1, k_2, k_3 =1}^L \frac{| u_L^n(1,k_3; 0 )|^2}
{\phi +2\sum_{i=1}^{2}[1-\cos (2\pi k_i /L_i)] +2[1-\cos (\pi (k_3
-1)/L_3)]}.
\label{sig}
\end{eqnarray}
In the limit of an infinite film geometry this equation yields ($L_3 = L$
is
kept finite)
\begin{equation}
\fl \lim_{L_1, L_2 \rightarrow \infty} \langle \sigma^2 (r_1 , r_2, 1
)\rangle
=\frac{1}{K} [ W_3 (\phi ) - 1/12 + \phi W_3 (\phi )/12 + W_2 (\phi )/2L ].
\label{ss}
\end{equation}
Hence, at the critical point $K=K_c$ of the infinite system ($L = \infty$),
by taking into account the bulk spherical constraint at $\phi_0 =0$, namely
$W_3 (0) = 2K_c$, one obtains that the mean square length $\rho_s$ of the
spins at the Neumann boundary of the standard shperical model equals
precisely the critical value  $\rho_c$ for the surface  spins in the
modified
spherical model.

Finally, we note that by taking
\begin{equation}
\omega =-8\pi \left( K_c-K\right) ,  \label{omn}
\end{equation}
one obtains for the considered system with layer geometry and
Neumann-Neumann
boundary conditions $\gamma _{1,1}=2$, which is the corresponding critical
exponent for the $O(n)$ models in the limit $n\rightarrow \infty $. In that
case $\gamma _{1,1}^{\prime }$ exists too, and $\gamma _{1,1}^{\prime }=1$.
Obviously, such a choice of $\omega $ defines an effective Hamiltonian that
leads to an exactly solvable model with the critical exponents stated
above.

Now we pass to the analysis of the behaviour of $\chi_{1,1}(T;\rho )$ given
by equation (\ref{hif}) where $\omega_0$ is determined as a function of $K$
and $\rho $ from equation (\ref{seq}).

\subsection{Critical behavior of the local surface susceptibility}

Here we confine our analysis to the surface critical regimes that emerge
on approaching the bulk critical temperature from above, i.e. when $K =
K_c + \Delta K$, where $\Delta K <0$ and $|\Delta K | \rightarrow 0$. Then,
as it is well known, the leading asymptotic form of the bulk spherical
field
follows from the asymptotic expansion
\begin{equation}
W_3 (\phi) = 2K_c - (4\pi )^{-1} \phi^{1/2} + O(\phi), \hspace{0.5cm} \phi
\downarrow 0 \label{w3}
\end{equation}
and reads \cite{Ba74}
\begin{equation}
\phi = 64  \pi^2 |\Delta K|^{2}, \hspace{0.5cm} \Delta K \uparrow 0 .
\label{phi}
\end{equation}

From expression (\ref{hif}) it is clear that the local surface
susceptibility
may exhibit divergent behaviour in two different regimes: (a) when
$\omega_0 \downarrow  0$, and (b) when $\omega_0 \downarrow \omega_1 (\phi
)
\uparrow 0$. As it is clear from equation (\ref{seq}) and the above
mentioned properties of the function $G_3 (\phi ,\omega)$, the first regime
may occur only when $K\rho \uparrow K_c \rho_c$, which, in view of our
assumption $\Delta K \uparrow 0$, requires $\rho = \rho_c$. The second
divergent regime of the local surface susceptibility takes place at any
fixed
$\rho > \rho_c$. Below we derive the leading-order asymptotic solutions for
$\omega_0 $ in each of the two cases.

{\em Case (a):} $\rho = \rho_c$. To obtain an asymptotic expansion of
$G_3 (\phi ,\omega)$ in both arguments $\phi \downarrow 0$ and
$\omega \downarrow 0$, we notice that the integral in the r.h.s of equation
(\ref{gpe}) diverges at $\phi = \omega =0$ and the divergence arises from
the integration over the neighbourhood of the point $\theta_1 =\theta_2
=0$.
Therefore, its leading-order asymptotic behaviour is given by the small
argument expansion of the trigonometric functions which yields
\begin{equation}
\fl
G_3^{+} (\phi , \omega ) = 2K_c \rho_c - (2\pi )^{-1} \phi^{1/2} +
(\omega /2\pi) \ln(\phi^{1/2}+\omega ) + O(\phi)+O(\omega).
\label{g3e}
\end{equation}
By setting $\rho = \rho_c$, we obtain that $\omega_0 \downarrow 0$ obeys
the
asymptotic equation ($\Delta K \uparrow 0$)
\begin{equation}
- (\omega_0 /2\pi)\ln (8\pi |\Delta K|+\omega_0) = |\Delta K|/(6 K_c).
\label{aeom}
\end{equation}
The solution which tends to zero from above as $|\Delta K| \rightarrow 0$
is
\begin{equation}
\omega_0 \simeq - \frac{\pi |\Delta K|}{3 K_c \ln (8\pi |\Delta K|)} .
\label{om0}
\end{equation}
Obviously, this critical regime leads to $\gamma_{1,1} = 1$.

{\em Case (b):} $\rho > \rho_c$. The asymptotic behaviour of $G_3 (\phi ,
\omega)$ as $\phi \downarrow 0$ and $\omega \uparrow 0$, so that $\omega >
\omega_1 (\phi)$, is readily obtained  from the exact representation
(\ref{gpm}) and the expansions (\ref{w2}) and (\ref{g3e}):
\begin{equation}
\fl
G_3^{-} (\phi , \omega ) = 2K_c \rho_c - (2\pi )^{-1} \phi^{1/2} -
(|\omega |/2\pi) \ln(\phi^{1/2} -|\omega |) + O(\phi)+O(\omega).
\label{g3m}
\end{equation}
At fixed $\Delta \rho >0$ the leading-order equation for the surface
spherical
field becomes
\begin{equation}
- (|\omega_0 |/2\pi)\ln (8\pi |\Delta K|- |\omega_0|) =2K_c \Delta \rho.
\label{aeo}
\end{equation}
Assuming $|\omega_0 | = 8\pi |\Delta K| - x$, where $x = o(|\Delta K|)$,
one
obtains
\begin{equation}
\omega_0 \simeq -8\pi |\Delta K| + \exp \left(-\frac{K_c \Delta \rho}{2
|\Delta K|}\right).
\label{o}
\end{equation}
Therefore, in this  critical regime the local surface susceptibility
diverges
exponentially as the bulk critical temperature is approached from above:
\begin{equation}
\chi_{1,1}(T;\rho ) \simeq \frac{1}{2K} \exp \left(\frac{K_c \Delta \rho}{2
|\Delta K|}\right) .
\label{chiex}
\end{equation}
This behaviour reminds the one of a two dimensional system close to $T=0$.
The
fact that the surface is coupled to an infinite three dimensional system
is reflected in the replacement of $T=0$ by the bulk critical temperature
$T=T_c$.

Finally, if $\rho<\rho_c$ it is easy to see that (\ref{seq}) has a finite
solution $\omega_0(\rho,K)$, where $0<\omega_0<1$, when $\Delta K \uparrow
0$.
The last actually follows from the inequalities
\begin{equation}
G_3 (0,\phi)>W_3 (\phi)>G_3 (1,\phi).
\label{uneq}
\end{equation}
Thus, if $\rho = 1$ the local surface susceptibility $\chi_{1,1}(T_c;1)$ is
finite.

\subsection{Critical behavior of the local susceptibility around the middle
of the system}

For the local susceptibility $\chi _{l,l}(T;\rho )$ from (\ref{ukw}) and (%
\ref{hild}) one obtain explicitly
\begin{eqnarray}
\fl \chi _{l,l}(T;\rho ) =\frac{1}{K}\lim_{L\rightarrow \infty }\frac{1}{L}
\sum_{k=1}^L\frac{\{ \sin [l\varphi_L^n(k;\omega )]-(1-\omega )\sin
[(l-1)\varphi _L^n(k;\omega )]\}^2}{\phi +2[1-\cos \varphi_L^n(k;\omega
)]}\times   \nonumber \\
\left\{ 1-2(1-\omega )\cos \varphi _L^n(k;\omega) +(1-\omega
)^2+[1-(1-\omega )^2]/L\right\}^{-1}.  \label{hl}
\end{eqnarray}
We will be interested only in the behavior of this quantity around the
middle of the system. Let us set $l=(L+1)/2.$ Then, from (\ref{hl}) it
follows
\begin{eqnarray}
\fl \chi _{l,l}(T;\rho ) =\frac 1K\lim_{L\rightarrow \infty }\frac
1L\sum_k\left[ \phi +2[1-\cos \varphi _L^n(k;\omega )]\right] ^{-1}\times
\nonumber  \\
\left\{ 1-\frac{1-(1-\omega )^2}{L\left[ 1-2(1-\omega )\cos \varphi
_L^n(k;\omega )+(1-\omega )^2\right] +1-(1-\omega )^2} \right\} ,
\label{hi2}
\end{eqnarray}
where the summation is  over the roots of equation (\ref{cos}) only. Having
in mind the properties of the roots $\varphi _L^n(k;\omega )$, it is easy
to
see that in the limit $L\rightarrow \infty $ this equation leads to
\begin{equation}
\chi_{\infty ,\infty }(T;\rho )=\frac{1}{2K}W_1 ( \phi_0)
\label{hin}
\end{equation}
for any $\omega $. We recall now that considering $\omega $ as a free
parameter, at $\omega =0$ one has the standard spherical model under
Neumann-Neumann boundary condition and at $\omega =1$ the corresponding one
with Dirichlet-Dirichlet boundary conditions. The above result shows that
the behavior of $\chi _{\infty ,\infty }(T;\rho )$ does not actually depend
on the boundary conditions. From the temperature dependence of $\phi _0$
around the bulk critical temperature  $\phi _0\simeq [ 8\pi (K_c-K)]^2$
\cite{BF} and the expansion of $W_1 ( \phi )$ for small values of the
argument
\cite{BF},
\begin{equation}
W_1 ( \phi ) =\frac{1}{2}\phi^{-1/2}+O(\phi ^{1/2}),  \label{W1}
\end{equation}
we obtain $\gamma _{\infty ,\infty }=1$. It is clear that the same will be
true for any layer at finite distance from the middle of the system. The
above result has been derived in \cite{Ba74} for a spherical model under
Dirichlet-Dirichlet boundary conditions (see Eq. (82) in \cite{Ba74}). Here
we simply show that it does not depend on the boundary conditions if they
are identical at both the boundaries: just due to the symmetry the system
models by itself an analogue of the Neumann boundary at the middle layers.

\section{Discussion}

In the present article the surface critical behaviour of a modified three
dimensional mean spherical model with a $L$-layer film geometry under
Neumann-Neumann boundary conditions is considered. The standard spherical
model
is modified in the sense that in addition to the usual bulk spherical
constraint a second spherical field is included in the Hamiltonian to fix
the
mean square value of the spins at the boundaries at some value $\rho > 0$.
We are interested mainly in the upper critical behaviour of the local
susceptibilities $\chi_{1,1}$ and $\chi_{l,l}$ with $l$ close to the
middle of the system. The surface susceptibility $\chi _{1,1}$ and the
local
susceptibility $\chi_{\infty,\infty}$ are evaluated exactly and the
corresponding results are given by equations (\ref{hif}) and (\ref{hin}),
respectively.

It is shown that the behaviour of $\chi _{1,1}(T;\rho )$ depends
crucially on $\rho$. At $\rho =1$ we find that $\chi_{1,1}$ is finite at
the bulk critical temperature $T_c$, in contrast with the recently
derived value $\gamma_{1,1}=1$ in the case of just one global spherical
constraint. The result $\gamma_{1,1}=1$ is recovered only if $\rho =\rho_c=
2-(12 K_c)^{-1}$, where $K_c$ is the dimensionless critical coupling. When
$\rho >\rho_c$, the local surface susceptibility $\chi_{1,1}$ diverges
exponentially as $T\rightarrow T_c^{+}$, see equation (\ref{chiex}). The
calculation of the mean square value $\rho _s$ of the spins at the Neumann
boundary in the standard spherical model elucidates the appearance of the
critical value of $\rho =\rho _c$: it turns out that at the bulk critical
point
$\rho _c=\rho _s$, see equation (\ref{ss}) at $K=K_c$, $\phi =0$ and
$L=\infty$. As it is expected, the behaviour of the local susceptibility
$\chi_{\infty,\infty}$ turns out to be independent of the boundary
conditions
if they are the same at both boundaries. Just due to the symmetry, the
system
models by itself an analogue of the Neumann boundary at the middle layers
which
leads to $\gamma _{\infty ,\infty }=1$ (see Section 3.2 for details).
By considereing the second spherical field as an independent free
parameter,
we rederive in an uniform way the previously known critical properties of
the
local surface susceptibility. They follow directly from equation
(\ref{hif})
at $\omega=0$, for Neumann-Neumann (see (\ref{hiN})), and $\omega=1$, for
Dirichlet-Dirichlet boundary conditions. For $\omega \neq 0,1$ equation
(\ref{hif}) gives the corresponding result for the so-called "intermediate"
\cite{GF90} boundary conditions. From these results one concludes that
$\chi_{1,1}$ diverges at $T=T_c$ only under Neumann boundary conditions.
Finally, an effective hamiltonian which leads to an exactly solvable model
with $\gamma _{1,1}=2$, the value for the $n\rightarrow \infty $ limit of
the
corresponding $O(n)$ model, is proposed. It is given by equation (\ref{H})
where one has to set $v=-8\pi K(K_c-K)$, see (\ref{omn}).

We emphasize that the spherical model under nonperiodic boundary conditions
is
not in the same surface universality class as the corresponding $O(n)$
model
in the limit $n\rightarrow \infty$, in contrast with the bulk universality
classes. For example $\Delta_{1}^{{\rm o}}=1$ and $\Delta_{1}^{{\rm sb}}=2$
for the $O(\infty)$ model, but $\Delta_{1}^{{\rm o}}=1/2$ and
$\Delta_{1}^{{\rm sb}}=3/2$ for the spherical model. The results presented
above show that the properties of the model are improved by introducing a
second spherical constraint in the sense that they are closer, in a certain
way, to the corresponding ones for the $O(n)$, $n\rightarrow \infty$,
model.
It seems clear that in order to obtain "correct" surface critical
properties,
one has to impose a separate spherical constraint on each layer paralell to
the surface.

\section*{Acknowledgment}

This work was supported by the Bulgarian National Foundation for Scientific
Research, grant MM-603/96.

\end{document}